\documentclass[reprint,
 amsmath,amssymb,
 aps,
pra,
]{revtex4-2}

\usepackage{graphicx}
\usepackage{dcolumn}
\usepackage{bm}
\usepackage{color}

\begin{document}

\preprint{APS/123-QED}

\title{AlF-AlF sticking time and prospects for ultracold dimers}

\author{Mahmoud A. E. Ibrahim$^{a,b}$}
\author{Mateo Londoño$^a$}
\author{Jes\'us P\'erez-R\'ios$^a$}%
 \email{jesus.perezrios@stonybrook.edu}
\affiliation{%
$^a$Department of Physics and Astronomy, Stony Brook University, Stony Brook, NY 11794, USA.\\
$^b$Department of Physics, Faculty of Science, Assiut University, Assiut 71515, Egypt.
}%

\date{\today}

\begin{abstract}
We report on the sticking time of the AlF dimer in the ultracold regime. We employ a full-dimensional potential energy surface for AlF–AlF, constructed using a machine learning approach [X. Liu et al., J. Chem. Phys. 159, 144103 (2023)], to compute the density of states using a semi-classical counting method. Next, using the Rice-Ramsperger-Kassel-Marcus (RRKM) theory, we determine a sticking time of 216.3 ns, which is shorter than that of other previously reported dimers. We explain these results in light of the ratio of the dissociation energy of the complex to the dissociation energy of the molecule, yielding a computationally inexpensive scheme to estimate the sticking time of collisional complexes.

\end{abstract}

\maketitle

\section{Introduction}

Applications and prospects of ultracold molecules in quantum technologies~\cite{Jaksch2018,Carr2009,reviewquantumsimulations}, quantum information science~\cite{pendular2013,Robust,Rabl2006,DeMille2002,computationdipolar}, fundamental physics~\cite{Safronova2018,Essig2019,Borkowski2019}, many-body physics~\cite{Micheli2006,JPR2010,Greiner2002,manybody,manybody2,quantum1}, and metrology~\cite{McGuyer2015,Wall2016,Kondov2019}, are contingent on two main assumptions: molecules can be brought to the ultracold regime, and ultracold molecules are chemically stable. In that regard, it has been shown that ultracold collisions of diatomic molecules form long-lived collision complexes in a phenomenon known as sticking, which in turn introduces loss mechanisms that limit the trap lifetime of ultracold molecules~\cite {mayle2012statistical,observation,observation_bis,observationHamburg,observationRbCs,ChristianenPRL,christianen2019quasiclassical,Bimolecular,observationBloch,Bause2023}. Therefore, it is necessary to characterize and understand the nature of sticky collisions, and more importantly, to calculate their lifetime.

Most of the theoretical efforts have been dedicated to studying bi-alkali systems due to their relevance in cold molecular sciences~\cite{Bimolecular}. However, more molecules can be brought to the ultracold regime using direct cooling techniques, i.e., employing a reservoir or an external field to remove the kinetic energy of the molecules, such as AlF~\cite{Truppe2019}, CaH~\cite{CaH,CaH2}, AlCl~\cite{AlCl}, to cite a few. Indeed, it has recently been possible to load AlF molecules into a magneto-optical trap~\cite{padillacastillo2025magnetoopticaltrappingaluminummonofluoride}, thereby expanding the family of ultracold molecules. These molecules are significantly different from bi-alkali molecules, especially AlCl and AlF, which are characterized as closed-shell and deeply bound molecules. No effort has been dedicated to studying the sticky collisions of molecules relevant to direct cooling techniques, except for the work on CaF-CaF~\cite{sardar2023sticking}. More importantly, the reason why some bi-alkali molecules show a long sticking time is not well understood. As a result, there is no intuitive model or general explanation for the physics behind sticky collisions. 
 
In this work, aided by a machine learning approach for the potential energy surface, we calculate the sticking time of AlF-AlF relevant for the stability of ultracold AlF molecules. We notice some key differences between AlF and bi-alkali molecules that we rationalize in terms of the potential energy landscape. In particular, we find that the ratio of the dissociation energy of the complex to the diatomic molecule correlates with the sticking time. These findings helped us to develop a qualitative model of the nature of sticky collisions that can help us anticipate the stability of ultracold polar molecules with minimal calculations, such as CaH, AlCl, and MgF.



\section{Methodology and computational details}

In this section, we describe our computational methods. In subsection \ref{a}, we give a brief account of the development of the six-dimensional AlF dimer PES using a machine learning approach embedded within an active learning scheme \cite{liu2023molecular}. In subsection \ref{b}, we describe the numerical details of the calculation of the density of states of the AlF dimer. Finally, in subsection \ref{c}, we detail the \textit{ab initio} calculations we performed for CaH, AlCl, and MgF.

\subsection{The AlF-AlF potential energy surface}\label{a}

We use a full-dimensional potential energy surface (PES) for the AlF dimer by training a machine learning model on \textit{ab initio} points using an active learning scheme, following Liu et al.~\cite{liu2023molecular}. This approach does not require specific long-range information, and it is highly efficient. For instance, for molecular dynamics calculations for AlF-AlF collisions, the methodology only requires $\lesssim 0.01\%$ of the configurations to be calculated \textit{ab initio}, which accelerates dynamics calculations without giving up the accuracy of the PES.

\begin{figure}[h]
\centering
 \includegraphics[width=\linewidth]{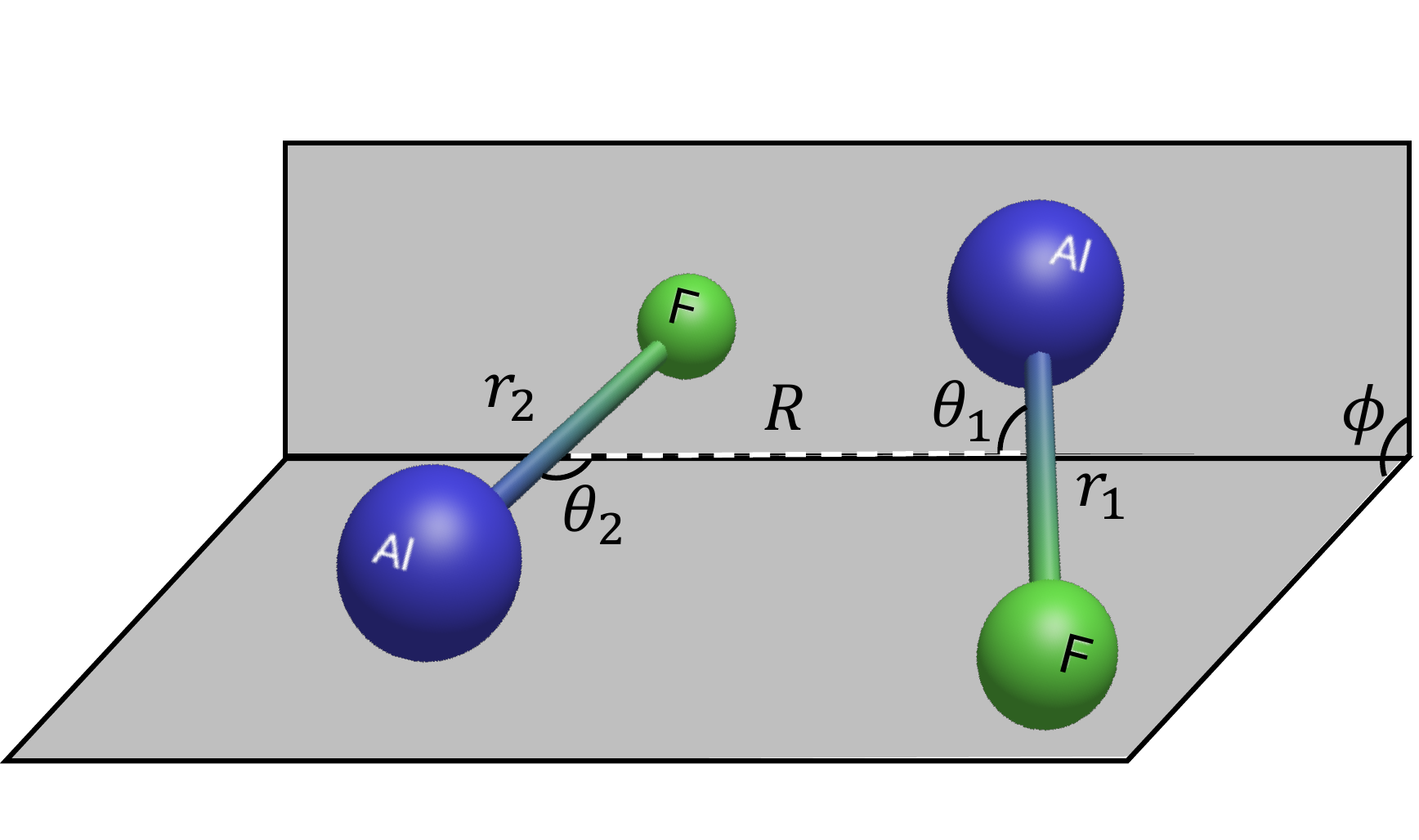}  
\caption{\label{fig:jacobi} Jacobi coordinates of AlF-AlF collision complex. $R$ is the intermolecular distance, $r_1$ and $r_2$ are the internuclear distances, $\theta_1$ and $\theta_2$ are the polar angles and $\phi$ is the torsion angle.}
\end{figure}

Diatomic dimers are better described in Jacobi coordinates, as displayed in Fig.~\ref{fig:jacobi}: $r_1$ and $r_2$ are the internuclear distances of the two colliding molecules; $R$, the intermolecular distance, is the length of the line connecting the centers of mass of the two molecules; $\theta_1$ is the polar angle between $\textbf{R}$ and $\textbf{r}_1$, and $\theta_2$ is the polar angle between $\textbf{R}$ and $\textbf{r}_2$. In this set of coordinates $\phi$ is the torsion angle formed between the $\textbf{R}$ - $\textbf{r}_1$ plane and $\textbf{R}$ and $\textbf{r}_2$. Similarly, $0\leq \theta_1,\theta_2 \leq \pi$ and $0 \leq \phi < 2\pi$. We let the origin of the body frame be the center of mass of the collision complex. We let $\textbf{R}$ and $\textbf{r}_1$ be on the \textit{xy} plane, with $\textbf{R}$ lying on the positive x-axis.


We assume that the energy of a given atomic configuration of the dimer $\textbf{x}$ is described by a Gaussian process as

\begin{equation}
    E(\textbf{x}) \sim \mathcal{G P} \left(m(f\left(\textbf{x}\right)),k(f\left(\textbf{x}\right),f\left(\textbf{x}'\right))\right).
\end{equation}
 where $\mathcal{G P} (.,.)$ is a Gaussian process specified by a mean function $m(.)$ and a covariance function (kernel) $k(.,.)$. The Jacobi coordinates represented by  $\textbf{x}$ only maintain translational and rotational symmetry but do not preserve the permutational invariance of the system. Thus, $f(\textbf{x})$ is introduced to maintain the permutational invariance of the system and is given by  
 \begin{equation}
     f(\textbf{x})=\hat{S}\{G(\textbf{x},\textbf{i})\}
 \end{equation}
where $\hat{S}$ the symmetrization operator accounts for the permutational symmetry and $\{G(\textbf{x},\textbf{i})\} $ is a series of n-body interaction terms that restore the permutational invariance of the molecular system where $\textbf{i}$ labels atoms within the system. The normalized interatomic distance $\bar{r}_{ij}=r_{ij}/r^*_{ij}$ is introduced to differentiate the chemical distinctions between various pairs of atoms, where $r^*_{ij}$ is the equilibrium interatomic distance between atoms labeled $i$ and $j$. To account for both Coulomb and long-range interactions, we considered the inverse interatomic distance $1/\bar{r}_{ij}$ as well as Morse-like exponential terms $e^{-\bar{r}_{ij}}$ to account for short-range effects.

The sampling of configurations was carried out by MD simulations, so that we could identify the most relevant configurations for the system at hand, which were selected as training data for our machine learning approach. Specifically, after running 8 trajectories calculated using ab initio molecular dynamics (AIMD) within the canonical ensemble, between 200 and 800 K, starting from different configurations, we sample configurations with intermolecular distance $R$ ranging from $\sim 1.5$ to 17.5~\AA, covering both short-range and long-range interactions. From these trajectories, we randomly choose 22390 ab initio configurations as the training set for our GPR. The details on the GPR algorithm can be found in Ref.~\cite{liu2023molecular}. In all these calculations, the ab initio forces were computed at the MP2 level of Møller–Plesset perturbation theory with the aug-cc-pVQZ basis set. In contrast, the energies were calculated at the coupled clusters with singles, doubles, and perturbative triples [CCSD(T)] level of theory.

 \subsection{Sticking time}
\label{b}
Within the Rice-Ramsperger-Kassel-Marcus (RRKM) formalism, the sticking time $\tau$ or the lifetime of the complex is given by~\cite{mayle2012statistical}
\begin{equation}
    \tau=\frac{h \rho}{N^{(0)}},
    \label{stick}
\end{equation}
where $\rho$ represents the density of states (DOS), $h$ is the Planck constant and $N^{(0)}$ is the number of outgoing channels. For the case at hand, AlF-AlF, $N^{(0)}=1$ since we consider both molecules in the rovibrational ground state, and hence no inelastic channels are available. In the absence of external fields, the total angular momentum \textit{J} and its projection on the quantization axis \textit{M}, are conserve quantities and hence, good quantum numbers. Then, the DOS for a given energy, $E$, can be calculated in a quasi-classical framework as~\cite{christianen2019quasiclassical}

\begin{equation}
\begin{split}      
    \rho^{\text{AlF}+\text{AlF}}(E) =\frac{g_{\textbf{\textit{N}}Jp} 4 \pi^6 m_{\text{Al}}^6m_{\text{F}}^6}{h^9(2m_{Al}+2m_{F})^3g_{\text{AlFAlF}}} \\ \times \int
{\frac{R^4r_1^4r_2^4\sin^2{\theta_1}\sin^2{\theta_2}}{\det{\mathcal{I}(\textbf{q})}\sqrt{\det{\mathcal{A}(\textbf{q})}}} [E-V(\textbf{q})]^2} d\textbf{q} 
\end{split}
\end{equation}
where $m_{\text{Al}}$ and $m_{\text{F}}$ are the masses of Al and F, respectively, and $q=(R,r_1,r_2,\theta_1,\theta_2,\phi)$ defines the atomic configuration given by the Jacobi coordinates of the collision complex. $g_{\textbf{\textit{N}}Jp}$ is the fraction of the classical phase space with quantum numbers \textbf{\textit{N}} and \textit{J} that is assigned with parity \textit{p}. Here, we use $g_{\textbf{\textit{N}}Jp}=1/2$. The degeneracy factor $g_{\text{AlFAlF}} = \Pi_i N_i !$  accounts for the indistinguishability of the monomers, where $N_i$ is the number of atoms per atomic species $i$; here $g_{\text{AlFAlF}}=4$. $V(\textbf{q})$ is the AlF-AlF PES. Details about the inertia tensor $\mathcal{I}(\textbf{q})$ and $\mathcal{A}(\textbf{q})$ can be found in Ref.~\citenum{christianen2019quasiclassical}.

For the computation of the DOS, we use a grid in \textit{R} of 60 points ranging from $1.4$~\AA\ to $\sim 10.0$~\AA; a grid of 32 points in $r_1$ and $r_2$ ranging from $0.4 $~\AA\ to $\sim 6.0$~\AA\ where we take $r_2 \geq r_1 $ and multiply the integral by a factor of 2. For $\theta_1$ and $\theta_2$, we used a 24-point Gauss-Legendre quadrature from 0 to $\pi$. Finally, we used an 8-point Gauss-Chebyshev quadrature for $\phi$ running from 0 to $\pi$ instead of $2 \pi$ and multiplied by another factor of 2, due to the symmetry of the dimer.

\subsection{Ab initio calculations}
\label{c}
We compare the dissociation energies of different collision complexes into two separate diatomic molecules to the dissociation energies of the diatomic molecules into two separate atoms. In particular, we consider the dissociation of X$_a$Y$_a$-X$_b$Y$_b$ complex into separate X$_a$Y$_a$ and X$_b$Y$_b$ diatomic molecules, where X is any of Al, Ca or Na and Y is any of H, F, Cl or K and the subscripts ($a$ and $b$) are labels to distinguish the atoms belonging to molecule $a$ from those belonging to molecule $b$. 

\begin{table}[h]
\begin{center}
\caption{Summary of ab initio methods (AIM) and basis sets used for different dimers, including the effective core potentials (ECP).}
\label{basis}
\begin{tabular}{lccccc}
 Complex&AIM method&Atom&ECP&Basis\\
\hline
\\
AlF-AlF$^a$&CCSD(T)& Al & - &aug-cc-pVQZ\\ 
\\
&& F & - &aug-cc-pVQZ\\ 
\\
AlCl-AlCl$^b$&CCSD(T)& Al & - & aug-cc-pVTZ \\ 
\\
&&Cl&ECP10MWB&aug-cc-pVTZ\\
\\
CaH-CaH$^b$&CCSD(T)&Ca&ECP10MDF&cc-pwCVTZ-PP\\
\\
&&H&-&aug-cc-pVTZ\\
\\
MgF-MgF$^b$&CCSD(T)&Mg&ECP10SDF&aug-cc-pVTZ\\
\\
&&F&-&aug-cc-pVTZ\\
\\
\\
$^a$ Ref. \cite{liu2023molecular}\\
$^b$ This work\\
\label{table1}
\end{tabular}
\end{center}
\end{table}

We performed geometry optimization to find the minima of the four-body potential energy surfaces for each of the AlCl-AlCl, CaH-CaH, and MgF-MgF dimers using the MOLPRO 2023.2 software package \cite{werner2020molpro}. We carried out geometry optimizations for all complexes at the CCSD(T) level of theory. For the CaH-CaH complex, we use Peterson's pseudopotential-based correlation-consistent polarized weighted core valence triple-$\zeta$ basis set for the Ca atom (cc-pwCVTZ-PP)~\cite{hill2017gaussian}, and model its 10 innermost electrons using Stuttgart/Koeln fully relativistic multielectron fit effective core potential (ECP10MDF)~\cite{lim2005all}. For the H atom, we use the correlation consistent polarized valence triple-$\zeta$ basis set with diffuse augmenting functions (aug-cc-pVTZ)~\cite{kendall1992electron}. In the case of the AlCl complex, we implement the aug-cc-pVTZ~\cite{kendall1992electron} basis set for both the Al and the Cl atoms, but we place a quasi-relativistic multielectron fit effective core potential (ECP10MWB)~\cite{lim2005all} on the Cl atom's 10 innermost electrons. Finally, for the MgF dimer, we use the aug-cc-pVTZ basis set and impose a relativistic single-electron fit effective core potential (ECP10SDF)~\cite{fuentealba1985pseudopotential} on Mg. For transition state (TS) search, we calculate the Hessian matrix using MP2. 

\section{Results and discussion}

\begin{figure}[h]
\centering
 \includegraphics[width=\linewidth]{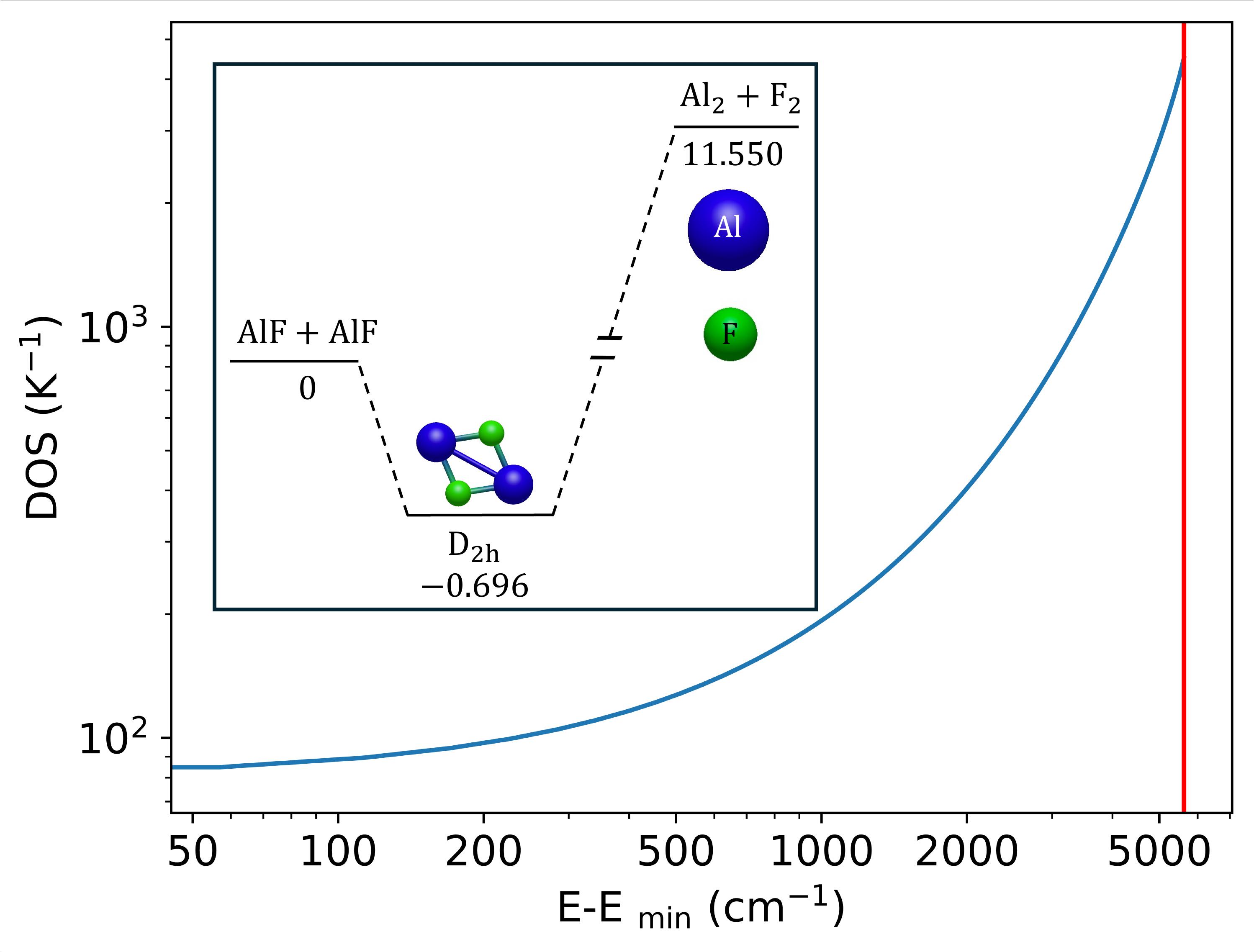}  
\caption{\label{fig:dos vs E} Density of states of the AlF-AlF complex as a function of energy on a double logarithmic scale. The vertical line indicates the dissociation energy of the complex into two AlF molecules. The inset diagrammatically shows the endothermic reaction AlF + AlF $\rightarrow$ Al$_2$ + F$_2$ on the singlet surface, the energies are in eV and details about the minimum configuration is listed in Table \ref{c}.}
\end{figure}

The DOS for AlF-AlF as a function of the energy referred from the global minimum of the PES $E-E_{\text{min}}$ is shown in Fig.~\ref{fig:dos vs E}. The global minimum of the AlF-AlF interaction potential $E_{min}=5613.62$~cm$^{-1}\equiv 0.696$~eV, is measured from the AlF-AlF asymptote and depicted by the dashed-red vertical line. A detailed scheme on the energy landscape for AlF-AlF collision on the singlet surface is shown in the inset of Fig.~\ref{fig:dos vs E}, emphasizing that the AlF + AlF $\rightarrow$ Al$_2$ + F$_2$ reaction is highly prohibited in the cold and ultracold regime due to an endothermic barrier of 11.550~eV. 

\begin{table*}
\small
\begin{center}
\caption{Optimized geometries of different dimers in terms of internuclear distances. Complex $D_o$ (eV) is the complex well depth from the separated diatomic molecules. X is any of Al, Ca, or Na, and Y is any of H, F, Cl, or K, and the subscripts ($a$ and $b$) are labels to distinguish the atoms belonging to molecule $a$ from those belonging to molecule $b$ in the asymptotic limit of the X$_a$Y$_a$-X$_b$Y$_b$ complex dissociation into separate X$_a$Y$_a$ and X$_b$Y$_b$ diatomic molecules. $\mathcal{R}$ is the dissociation energy of the four-body complex to the diatomic dissociation energy ratio.}
\label{c}
\begin{tabular*}{\textwidth}{@{\extracolsep{\fill}}lllcccccccc}
 Complex&\begin{tabular}{@{}cc@{}}
                 Surface
                 \end{tabular} &\begin{tabular}{@{}cc@{}}
                 Geometry
                 \end{tabular}&\begin{tabular}{@{}cc@{}}
                 \\X$_a$-Y$_a$ \\ (\AA)\\
                 \end{tabular}&\begin{tabular}{@{}cc@{}}
                 \\X$_b$-Y$_b$ \\(\AA)\\
                 \end{tabular}&\begin{tabular}{@{}cc@{}}
                 \\X$_a$-X$_b$ \\ (\AA)\\
                 \end{tabular}&\begin{tabular}{@{}cc@{}}
                 \\Y$_a$-Y$_b$\\ (\AA)\\
                 \end{tabular}&\begin{tabular}{@{}cc@{}}
                 \\X$_a$-Y$_b$\\ (\AA)\\
                 \end{tabular}&\begin{tabular}{@{}cc@{}}\\X$_b$ -Y$_a$\\(\AA)\end{tabular}&\begin{tabular}{@{}cc@{}}\\Complex\\ $D_o$ (eV)\end{tabular}&\begin{tabular}{@{}cc@{}}
                 $\mathcal{R}$
                 \end{tabular}\\ 
                 \\
                 
\hline
\\
$^a$AlF-AlF& Singlet & D$_{2h}$ & 1.8872 & 1.8872 &2.3212&2.3212&1.8872&1.8872&0.696&0.10\\ 
\\
$^b$AlCl-AlCl & Singlet   & D$_{2h}$ & 2.443 & 2.443 &3.624&3.277& 2.443 & 2.443&0.377&0.07\\ 
\\
$^b$CaH-CaH & Triplet  & D$_{2h}$ &2.184 & 2.184 & 3.497
 &2.619&2.184 & 2.184&1.801&1.06\\
\\
&Triplet  &C$_s$&2.128&2.042&3.715&3.838&5.672&2.208&0.982&0.58\\
\\
& Singlet  &D$_{2h}$& 2.194 & 2.194 & 3.506
 &2.641& 2.194 & 2.194&1.893&1.11\\
\\
& Singlet  &C$_s$& 2.280&2.031&3.482&3.915&5.513&2.137&1.525&0.90\\
\\
$^b$MgF-MgF & Triplet  & D$_{2h}$ &1,915&1.915&2.890&2.513&1.915&1.915&2.231&0.47\\
\\
& Triplet  & C$_s$ & 1.858&1.761&3.142&3.270&4.825&1.915&0.987&0.21\\
\\
& Singlet  & D$_{2h}$ & 1.912 & 1.912 & 2.881 &2.515& 1.912 & 1.912 &2.377&0.50\\
\\
& Singlet  & C$_s$ & 1.747 & 1.953 & 2.853 & 3.464 & 1.852 & 4.580 &2.143&0.45\\
\\
$^c$CaF-CaF & Triplet  & D$_{2h}$ & 2.151 & 2.151 & 3.388 &2.653&2.151 & 2.151&2.224&0.41\\
\\
&Triplet  & C$_s$ & 1.992 & 2.090 &3.723&3.780&5.624&2.180&1.850&0.33\\
\\
& Singlet  & D$_{2h}$ & 2.147 & 2.147 & 3.366 &2.667& 2.147 & 2.147&2.346&0.43\\
\\
& Singlet  & C$_s$ & 1.988 & 2.120 & 3.451 & 3.883 & 5.434 & 2.157&1.905& 0.35\\
\\
$^d$NaK-NaK &    &  D$_{2h}$ & 3.863 & 3.863 & 3.969 & 3.969 & 3.863 & 3.863&0.562&0.89\\
\\
&&C$_s$&3.969& 3.916 & 3.493 & 4.392 & 6.988 & 3.545&0.425& 0.67 \\
\\
\\
$^a$ Ref. \cite{liu2023molecular}\\
$^b$ This work\\
$^c$ Ref. \cite{sardar2023four,sardar2023sticking}\\
$^d$ Ref. \cite{christianen2019six}
\label{table:test}
\end{tabular*}
\end{center}
\end{table*}

Using Eq.~(\ref{stick}), the value of the DOS at the dissociation energy of the complex ($E-E_{\text{min}}=0$), we find that the sticking time is 216.3~ns, shorter than any previously reported dimer lifetimes in the ultracold regime. Why is that? The answer should focus on the peculiarities of the reactants and the properties of the four-body complex. In this line, we notice that AlF-AlF shows a single minimum while bi-alkali dimers show 2 minima (at least), and the reaction AlF + AlF is highly endothermic in comparison with bi-alkali systems. Indeed, most of the bi-alkali dimer reactions are either exothermic or slightly endothermic. Moreover, and more relevant is the fact that AlF molecules are deeply bound ($\sim 7$~eV) in comparison with bi-alkali molecules ($\sim 1$~eV) due to the different bonding mechanism in AlF (ionic-covalent) versus bi-alkali (Van der Waals). On the contrary, the dissociation energy of the four-body complex for bi-alkali systems and AlF is very similar, $\sim 1$~eV. Hence, the most striking difference between AlF and bi-alkali seems to be the dissociation energy of the four-body complex to the diatomic dissociation energy ratio, $\mathcal{R}$. 


\begin{figure*}
\centering
 \includegraphics[width=\linewidth]{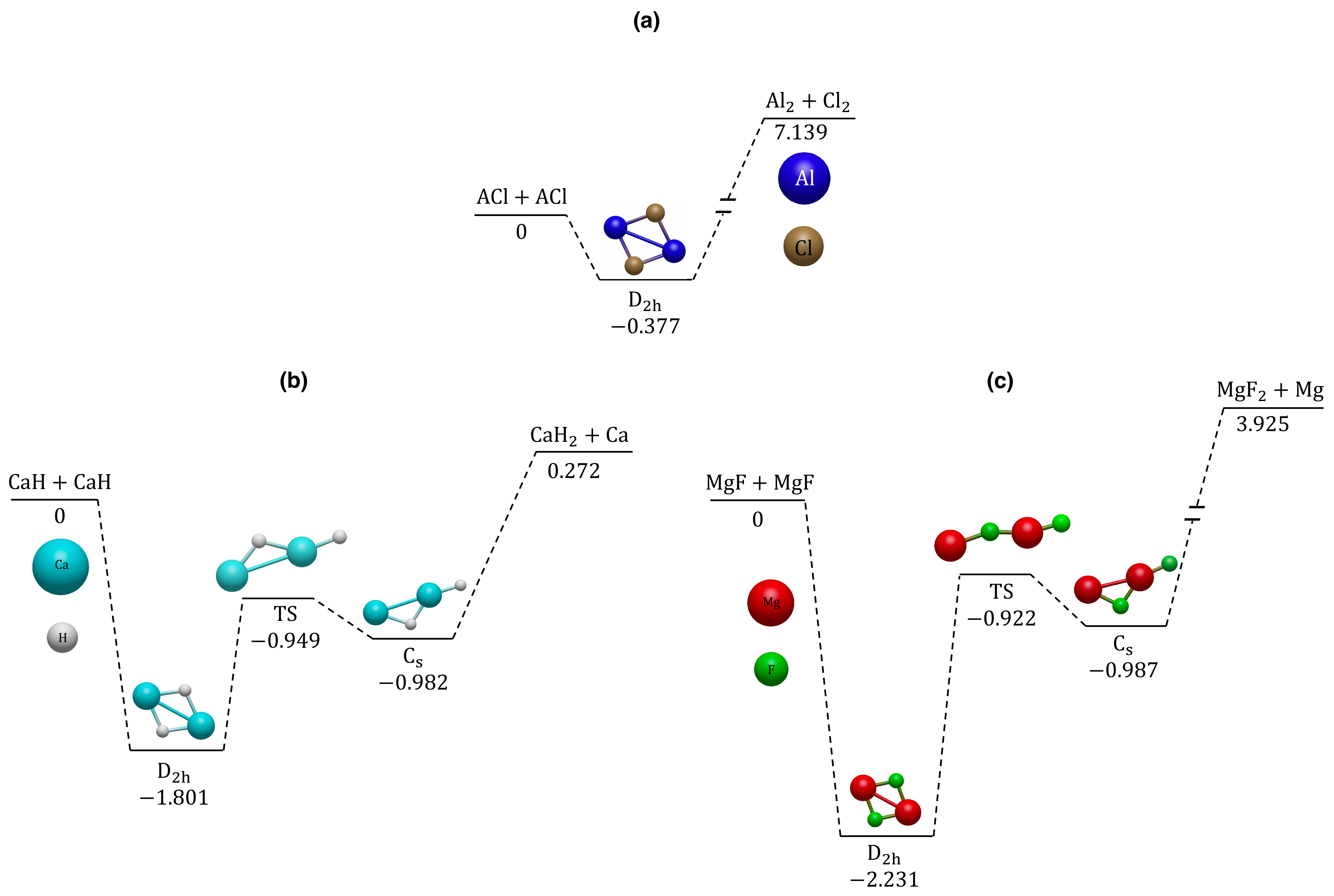}  
\caption{\label{fig:energy diagram} Reaction pathways of (a) AlCl + AlCl $\rightarrow$ Al$_2$+ Cl$_2$ on the singlet surface, and (b) CaH + CaH $\rightarrow$ CaH$_2$ + H and (c) MgF + MgF $\rightarrow$ MgF$_2$ + F on the triplet surface. The energies are in eV. The geometries of the global and local minima are listed in Table \ref{c}. The internuclear distances of the CaH dimer's transition state (TS) in panel (b) are given by X$_a$-Y$_a$=2.142~\AA, X$_b$-Y$_b$=2.019~\AA, X$_a$-X$_b$=3.718~\AA, Y$_a$-Y$_b$=4.018~\AA, X$_a$-Y$_b$=5.652~\AA,~ and  X$_b$-Y$_a$=2.297~\AA.~ The internuclear distances of the MgF dimer's transition state (TS) in panel (c) are given by X$_a$-Y$_a$=1.977~\AA, X$_b$-Y$_b$=1.688~\AA, X$_a$-X$_b$=3.687~\AA, Y$_a$-Y$_b$=3.425~\AA, X$_a$-Y$_b$= 5.368~\AA, ~and  X$_b$-Y$_a$=1.778~\AA. The X's denote Ca atoms in (b) and Mg atoms in (c), and the Y's denote H atoms in (b) and F atoms in (c) and the subscripts ($a$ and $b$) are labels to distinguish the atoms belonging to molecule $a$ from those belonging to molecule $b$ in the asymptotic limit of the X$_a$Y$_a$-X$_b$Y$_b$ complex dissociation into separate X$_a$Y$_a$ and X$_b$Y$_b$ diatomic molecules.}
\end{figure*}

Next, we explore $\mathcal{R}$ as the figure of merit for the estimation of the sticking time in ultracold dimers. To do that, we optimized the geometry of AlCl-AlCl, MgF-MgF and CaH-CaH--relevant molecules for cold molecular sciences, and the results, in comparison with CaF-CaF and NaK-NaK (as a representative case of bi-alkali molecule), are summarized in Table \ref{c} and displayed diagrammatically in Fig.~\ref{fig:energy diagram}. The table contains the optimized geometries, dissociation energies of the four-body complex, and $\mathcal{R}$. For all configurations, the internuclear distances stretched in comparison to the equilibrium internuclear distance $r_e$ of their diatomic monomers. We notice that $\mathcal{R}$ for AlF is smaller than in bi-alkali dimers, as anticipated. More surprising, is to realize that the $\mathcal{R}$ for AlY (Y is a hologen atom) is also smaller than XF and XH dimers, where X is an alkaline earth atom, yielding a hierarchy $\mathcal{R}_{\text{bi-akalki}} \gtrsim \mathcal{R}_{\text{XF}}>\mathcal{R}_{\text{AlY}}$, which correlates with the one in sticking times $\tau_{\text{bi-akalki}} \gtrsim \tau_{\text{XF}}>\tau_{\text{AlY}}$. Hence, we predict that the AlCl dimer will exhibit an even shorter sticking time than the AlF dimer; however, the MgF dimer is expected to be similar to the CaF dimer.

On the other hand, as shown in Fig.~\ref{fig:energy diagram}, it is clear that AlF-AlF and AlCl-AlCl are highly endothermic compared with bi-alkali or X-F and X-H reactions, where X is an alkaline-earth atom. Similarly, we notice that AlF and AlCl dimers show only one single minimum with D$_{2h}$ symmetry. On the contrary, the rest of the molecules (radicals and bi-alkali) present two minima, one with C$_s$ and another with D$_{2h}$ symmetry, with a transition state (TS) linking the two minima. The absence of a transition state linking two minima on the PES of AlF and AlCl dimers suggests that a trajectory will spend less time in the complex region, in comparison with other dimers where a transition state is present. Hence, in this way, explaining why AlF shows a short sticky lifetime. All these correlations point to a more profound difference that, we believe, is related to the binding mechanism of the dimer versus the diatomic molecule. In the case of bi-alkali molecules, the diatomic molecule exhibits a Van der Waals type of bonding, whereas the dimer displays predominantly covalent bonding. On the contrary, for AlY molecules, the binding mechanism for the molecule is mainly covalent with some ionic character to it, whereas the dimer shows Van der Waals bonding.

\section{Conclusion}

In this work, we report the lifetime of the AlF dimer in the ultracold regime, based on a machine learning potential energy surface for the dimer that includes all degrees of freedom. The density of states is $4.5 \times 10^{-3} \mu$K$^{-1}$, and its sticking time is 216 ns, which is $\lesssim 100$ times that of other theoretically studied dimers. We find that this discrepancy is related to the ratio of the dissociation energy of the complex to the dissociation energy of the parent molecule, indicating that this ratio correlates with the sticking time of dimers in the ultracold regime. We provide a physical model to predict the sticking time of ultracold dimers without computing the entire potential energy surface and without performing any state counting. Based on this model, we speculate that molecules deeply bound through covalent and ionic bonding character form Van der Waals four-body complexes, showing a smaller density of states, and accordingly, a shorter sticking time. Finally, thanks to our model, it is possible to screen stable molecules in the ultracold regime with minimal theoretical effort.

\begin{acknowledgments}
The authors acknowledge the support of the United States Air Force Office of Scientific Research [grant number FA9550-23-1-0202]. 
\end{acknowledgments}

\nocite{*}

\bibliographystyle{apsrev}
\bibliography{biblio}

\end{document}